\documentclass{ws-procs975x65-arxiv}            
\begin{document}
\title{\vspace{-0.5cm}
HOW FAR CAN A PRAGMATIST GO INTO QUANTUM THEORY?\\
A CRITICAL VIEW OF OUR CURRENT UNDERSTANDING OF QUANTUM PHENOMENA}

\author{A. S. Sanz}

\address{Instituto de F{\'\i}sica Fundamental (IFF-CSIC),\\
28006 Madrid, Spain\\
E-mail: asanz@iff.csic.es\\
http://fama.iff.csic.es/personas/asanz/}

\begin{abstract}
To date, quantum mechanics has proven to be our most successful
theoretical model.
However, it is still surrounded by a ``mysterious halo'' that
can be summarized in a simple but challenging question: Why quantum
phenomena are not understood under the same logic as classical ones?
Although this is an open question (probably without an answer), from
a pragmatist's point of view there is still room enough to further
explore the quantum world, marveling ourselves with new physical
insights.
We just need to look back in the historical evolution of the quantum
theory and thoroughly reconsider three key issues: (1) how this has
developed since its early stages at a conceptual level, (2) what kind
of experiments can be performed at present in a laboratory, and
(3) what nonstandard conceptual models are available to extract some
extra information.
This contribution is aimed at providing some answers (and, perhaps,
also raising some issues) to these questions through one of such
models, namely Bohmian mechanics, a hydrodynamic formulation of the
quantum theory, which is currently trying to open new pathways of
understanding.\footnote{It is a difficult task to capture and provide
within a few pages feelings and thoughts about the quantum theory
accumulated along years of work with analytical developments and
numerical simulations.
Hence this contribution only intends to be a modest conclusion from
such a work, which is, of course, open to debate.}
Specifically, the Chapter constitutes a brief and personal overview on
the historic and contextual evolution of this quantum formulation, its
physical meaning and interest (leaving aside metaphysical issues), and
how it may help to overcome some preconceived paradoxical aspects of
the quantum theory.
\end{abstract}


\bodymatter


\section{How Do We Understand the Quantum World?}
\label{sec1}

In 2011, Physics World honored Aephraim Steinberg and
colleagues,\cite{physworld2011} from the University of Toronto,
with the top one position in its yearly breakthrough ranking.
So far the quest for fundamental particles involves large energy and
length scales ---from Giga to Teraelectronvolts (e.g., 7~TeV in p-p
collisions), from tenths of meters (e.g., the LHC detector has about
14.6~m of diameter and 21.6~m of total length) to Kilometers (e.g.,
the LHC ring has about 27~Km length).
In contrast, Steinberg's team developed a very interesting top-table
experiment\cite{steinberg2011} at much smaller scales ---about 1.3~eV
(943~nm) and a few meters---, designed to shed some light on another
fundamental aspect of the quantum theory: elucidating how quantum
particles travel (on average) through physical space.
With the aid of the so-called ``weak measurement''
technique,\cite{aharonov1988} consisting of weakly perturbing the
quantum system at some time during its evolution and previous to its
definite measurement, this team has become ``the first to track the
average path of single photons passing through a Young's double slit
experiment ---something that Steinberg says physicists had been
``brainwashed'' into thinking is impossible.''
But, why is this experiment relevant at all?
The mathematical formulation of the quantum theory is neat and accurate,
and its applications have proven to be very powerful ---an important
``bite'' of the gross internal product in industrialized countries
relies on quantum mechanics, including developments and applications
in technological, energetic or biomedical areas.
From the electron to the Higgs, the success of quantum mechanics is
indisputable.
However, what do we really know about quantum systems?
The famous Bohr-Einstein debates\cite{bohr-einstein} ended in the
1930s with the orthodox or Copenhagian view of quantum systems, which
has healthfully survived to date: the quantum world is essentially
probabilistic and hence it does not make any sense asking questions
intended for going beyond probabilities.
Actually, these probabilities are such that if we try to determine
accurate (probabilistic) information about one of the variables ($A$)
from a pair of (classical) canonically conjugate variables, we will be
unable to obtain any relevant information about the other ($B$), and
vice versa.
This is the essence of Bohr's complementarity principle, which formally
translates into the well-known Heisenberg uncertainty relation,
\begin{equation}
 \Delta A \cdot \Delta B \ge \frac{\hbar}{2} ,
\end{equation}
where $\Delta$ denotes the dispersion in the measurement of either $A$
or $B$.

At a more pictorial level, complementarity is often regarded to
the alleged wave-corpuscle duality exhibited by quantum systems: an
electron behaves as a wave when it passes through a pinhole, and as
a corpuscle when it is acted by a highly energetic photon (e.g.,
$\gamma$ rays in Compton scattering).
Hence, in Young's two-slit experiment, if the electron passes without
ever being disturbed during the transit, we will observe a nice fringed
pattern: the electron distributes within some regions, while avoids
some others.
All about the electron distribution is known, but nothing about which
slit the electron passed through or, in other words, which momentum the
electron carried.
Trying to determine the latter would eventually lead to fringe erasure
because of the impossibility to measure both at the same time.
According to von Neumann,\cite{vonNeumann} this is a manifestation of a
non-unitary evolution process known as the ``collapse'' of the wave
function.
That is, while the system is unperturbed, it displays a unitary
(probability preserving) evolution in compliance with Schr\"odinger's
equation; once a measurement is performed, such unitarity is broken and
the system state randomly ``collapses'' onto any of the pointer states
of the measurement device with the probability prescribed by the system
wave function.
When this argumentation was proposed, it found the strong opposition of
Einstein and others, who thought that there should be something else, a
set of internal or {\it hidden} variables that would determine the
outcome, thus removing any trace of randomness.
For those physicists seeking for an objective or realistic description
of quantum systems, the wave function could not be complete because
it was not able to specifically determine all possible information
about the system.

It was a hard and thorny way to disprove von Neumann's statement,
according to which quantum mechanics does not admit hidden variables.
However, in 1952 David Bohm proposed\cite{bohm1952} a counterexample to
this theorem, showing that an account of the individual evolution of
the system is still compatible with the quantum theory by simply
assuming that the wave function acts like a field ``guiding'' the
system (a similar idea was proposed about 25 years earlier by Louis de
Broglie\cite{debroglie1927}).
As a consequence, there is no need for a ``collapse'' postulate ---nor
even the presence of an external observer.
This does not mean that the mathematical structure itself of quantum
mechanics has to be changed, but only that there is still much more
room for thinking quantum phenomena in different alternative ways.
Bohm's suggestion remained almost forgotten until the 1960s, when it
called the attention of John Bell.\cite{bell-bk}
While working at CERN, he decided to go back to von Neumann's
theorem and re-examine it with the purpose of systematizing which
properties should be satisfied by a quantum hidden-variable theory
in order to be valid.
In so doing, Bell noticed\cite{bell1964} that what makes quantum
mechanics so special is a feature lacking in classical mechanics,
namely nonlocality, i.e., the fact that any local disturbance in a
quantum system immediately affects the whole system.
The most striking example where this property manifests is in quantum
entanglement: the correlations between two spatially separate systems
that interacted in the far past are very important, since they can be
used to transfer quantum information between two distant places
without incurring in surperluminal signaling.

Bell's contribution not only started the revolution of quantum
technologies, leading to the development of the quantum information
theory, quantum computing, quantum cryptography, quantum teleportation,
etc., but indirectly he also motivated a reconsideration of Bohm's
approach.
Nowadays, for some people this is just an alternative quantum theory
(this being based on the ontology generated from it); for others it is
only an alternative formulation of the quantum theory.
If this is simply regarded as a matter of taste, and we decide to
remain at a more pragmatic level, what really matters is the fact that
Bohmian mechanics has open an alternative pathway to understand quantum
systems, justifying Steinberg's statement that our ``brainwashed''
view of quantum phenomena may be changed by experiments like the one
that he and his colleagues performed just on an optical bench.


\section{A Single-Event Prescription to Think Quantum Phenomena}
\label{sec2}


\subsection{Single-event experiments}
\label{sec21}

During the early days of quantum mechanics the weight of statistical
mechanics and thermodynamics was too strong ---Boltzman's shadow was
too long.
It is not strange, therefore, that the Copenhagen interpretation became
widely accepted (aided by the neopositivist thought-streams of those
years), healthfully surviving up to now.
However, at present quantum experiments can be performed in a, by far,
more refined way than in the first decades of the XXth century, thus
limiting the importance of statistics.
More specifically, the system evolution can be monitored in real time
(e.g., molecular configurations, entanglement dynamics, electron
ionization, surface diffusion, etc.), in contrast with traditional
spectroscopic measurements based on the energy domain.\footnote{Even
if from a purist's viewpoint this tracking is not exactly in time, the
fact is that we can reconstruct a whole ``movie'' of the system
time-evolution, something technically forbidden until recent times.}
Furthermore, since the 1970s we also have interesting interference
experiments corroborating that, even if we know absolutely nothing
about how each (quantum) particle behaves individually, at least we
know that it reaches the detector as a single, localized event, and
not as an extended wave.
From the former experiments with electrons\cite{pozzi,tonomura} to
the latter with large molecular complexes,\cite{arndt} it has been
confirmed that, as time proceeds and the accumulation of (randomly
distributed) particle arrivals on a distant detector becomes larger,
an incipient interference fringed pattern starts emerging from a
seemingly disordered distribution of single, (time) uncorrelated
detections, in clear correspondence with the outcome expected from
Schr\"odinger's equation.
Statistics thus plays an important role in accounting for the general
picture, but says nothing about how each individual detection takes
place.

Chapter~2 of Feynman's renowned {\it Lectures on Physics} starts
as\cite{feynman-bk3} ``In this chapter we shall tackle immediately the
basic element of the mysterious behavior in its most strange form.
We choose to examine a phenomenon which is impossible, {\it absolutely}
impossible, to explain in any classical way, and which has in it the
heart of quantum mechanics.
In reality, it contains the {\it only} mystery.
We cannot make the mystery go away by ``explaining'' how it works.
We will just tell you how it works.
In telling you how it works we will have told you about the basic
peculiarities of all quantum mechanics.''
Effectively, the two-slit experiment summarizes the essence
(``mystery'') of quantum mechanics.
However, like Dirac, Feynman also thought that each individual
particle self-interfered, apparently being unaware of Pozzi's and
Tonomura's experiments on the two-slit experiment.
His scientific authority reinforced the Copenhagian viewpoint, but
should things have been different if Feynman would have watched
Tonomura's movie\cite{tonomura} of his two-slit experiment with
electrons?
Well, although evidently we have no answer to this question, perhaps
we could say that, for someone who succeeded in introducing the concept
of trajectory into quantum mechanics (in spite of Bohr's disapproval),
things would have been, at least, a bit different.

In any case, it is important to keep in mind three very simple ideas:
\begin{itemize}
\item[i)]
{Appealing to Occam's razor, there is no reason to think that the wave
function collapses to a local point upon detection, as formerly stated
by von Neumann.\cite{vonNeumann}}
\item[ii)]
{Experiments corroborate a statistical origin of interference patterns
(and, in general, any quantum trait), consisting of a large
accumulation of single, localized events.\cite{pozzi,tonomura,arndt}}
\item[iii)]
{Even if all particles are generated at the same source (where they can
interact), once they are released experiments also confirm that each
particle arrival is independent of all other previous or subsequent
ones (see more recent experiment in \cite{pozzi}).
That is, one particle is totally unaware of what others do.}
\end{itemize}
Accordingly, one cannot avoid thinking whether single-event
descriptions are affordable in quantum mechanics, even being aware that
the evolution of a single particle cannot be tracked (at least, not
with the current technology) without irreversibly perturbing it.
In some sense, it would still be possible to work at the level of
classical fluids: we know nothing about the individual motion of the
fluid's constituents, but still its collective dynamical properties
could be determined by means of a streamline analysis.


\subsection{Bohmian mechanics}
\label{sec22}

For simplicity, let us consider the nonrelativistic Schr\"odinger
equation,
\begin{equation}
 i\hbar\frac{\partial \Psi}{\partial t} =
 \left( -\frac{\hbar^2}{2m}\nabla^2 + V \right) \Psi ,
\end{equation}
which describes how the wave amplitude $\Psi$ associated with a
physical system of mass $m$ evolves in time through a given
configuration space accounting, for instance, for the system position.
Statistical information about the possible outcomes that can be
expected at any place and time are determined from the probability
density $|\Psi|^2$.
It is at this point where Bohm\cite{bohm1952} felt the need to
introduce the concept of hidden variable, just as a way to test the
validity of the assumption that ``the most complete possible
specification of an individual system is in terms of a wave function
that determines only probable results of actual measurement processes.''
These hidden variables in principle would ``determine the precise
behavior of an individual system, but which are in practice averaged
over in measurements of the types that can now be carried out.''
The evolution of these hidden variables, which he identifies with
individual realizations (trajectories) of the system, arises after
recasting Schr\"odinger's equation as a set of two real equations
of motion, one for the probability density $\rho$ and another one for
the phase $S$ of the wave function (when the latter is expressed in
polar form, $\Psi = \rho^{1/2} e^{iS/\hbar}$).
This gives rise to the continuity equation and a Hamilton-Jacobi-like
equation, from which Bohm postulates that the system trajectories
evolve according to the equation of motion
\begin{equation}
 {\bf v} = \dot{\bf r} = \frac{\nabla S}{m} ,
\end{equation}
where ${\bf v}$ is a local velocity field.
Within this scenario, quantum systems thus consist of a wave field
($\Psi$) and a particle guided by this field, which follows a
trajectory ${\bf r}(t)$ in configuration space.
This is essentially what is nowadays known as Bohmian
mechanics,\cite{sanz-bk} although the more philosophical (ontological)
aspects are or are not seriously considered depending on the author.

Leaving aside metaphysical connotations, from a pragmatist's viewpoint,
Bohmian mechanics is basically a fluid-dynamical description for
quantum systems.
It is worth stressing that the same ideas were already proposed
(although not in the context of hidden variables) by de
Broglie\cite{debroglie1926} and Madelung,\cite{madelung1927} and
contemporarily to Bohm by Takabayasi.\cite{takabayasi}
Furthermore, it should be clear that this formulation is totally
equivalent and is at the same level as any other of the more
traditional ones (Schr\"odinger's, Heisenberg's, Dirac's, Moyal-Wigner,
etc.).
Each formulation emphasizes a different aspect of the quantum theory;
Bohmian mechanics stresses the fact that quantum systems can be
associated with a quantum fluid (not in vain Schr\"odinger's equation
is just a diffusion equation with an imaginary diffusion constant,
$i\hbar/2m$).
Actually, descriptions like the Bohmian one in terms of streamlines
are rather common in the literature,\cite{sanz2014} and in recent
years it has been possible to recreate Bohmian-like systems (in fact,
deBroglian ones) by means of classical fluid-dynamical
experiments.\cite{couder2006,bush2013}

\begin{figure}
 \begin{center}
 \includegraphics[width=10cm]{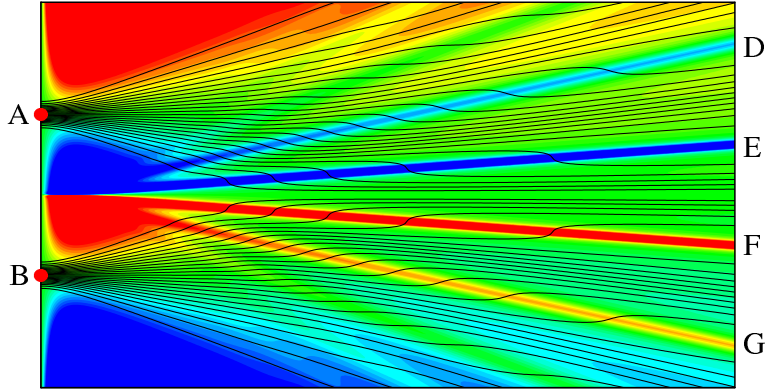}
 \caption{Numerical simulation of Young's two-slit experiment:
  the slits centered at A and B, along the $x$ direction, produce
  two Gaussian wave packets, which eventually superimpose and generate
  in the long time (Fraunhofer) regime\cite{sanz2012} a series of
  interference fringes (D, E, F, G).
  Black solid lines represent ensembles of Bohmian trajectories leaving
  each slit.
  The background contour-plot corresponds to the velocity field as
  time proceeds (time increases along the horizontal axis): from blue
  to red, increasing value of the velocity from negative to positive,
  respectively.
  Note that the non-crossing appears as a consequence of the sudden
  change of sign undergone by the velocity field along the system
  symmetry line ($x=0$).}
 \label{fig1}
 \end{center}
\end{figure}

Rather than hidden variables, Bohmian mechanics constitutes a valuable
tool that allows us to determine ``hidden'' quantum properties, i.e.,
properties that are not evident within other formulations of the
quantum mechanics.
For example, based on the Bohmian non-crossing property,\cite{sanz2008}
i.e., that Bohmian trajectories cannot pass through the same point of
the configuration space at the same time, we find that quantum flows
do not mix in configuration space, as Steinberg's experiment
shows,\cite{steinberg2011} and accordingly one can properly specify
tubes along which the probability remains constant at any
time.\cite{sanz2013}
Also, based on the non-crossing property, one notices that the physical
implications of the superposition principle go further beyond from its
simple mathematical application due to the phase dynamics involved,
which immediately modifies the velocity map throughout the whole
configuration space.\cite{sanz2008,sanz2012}
In any case, it is worth stressing that Bohmian trajectories only
provide us with hydrodynamical information of the quantum fluid or wave
field itself, but not of what is in there behind it ---that is, how
the average ensemble evolves, but not how a real individual particle
moves.\cite{sanz2014,sanz2012}
A simple example of Bohmian trajectories illustrating the renowned
Young two-slit experiment is displayed in Fig.~\ref{fig1}, where the
background contour-plot represents the nonlocal velocity field
pervading the whole configuration space.

\section{Wheeler's Delayed Choice Experiment Revisited}
\label{sec3}

\begin{figure}[t]
 \begin{center}
 \includegraphics[width=6cm]{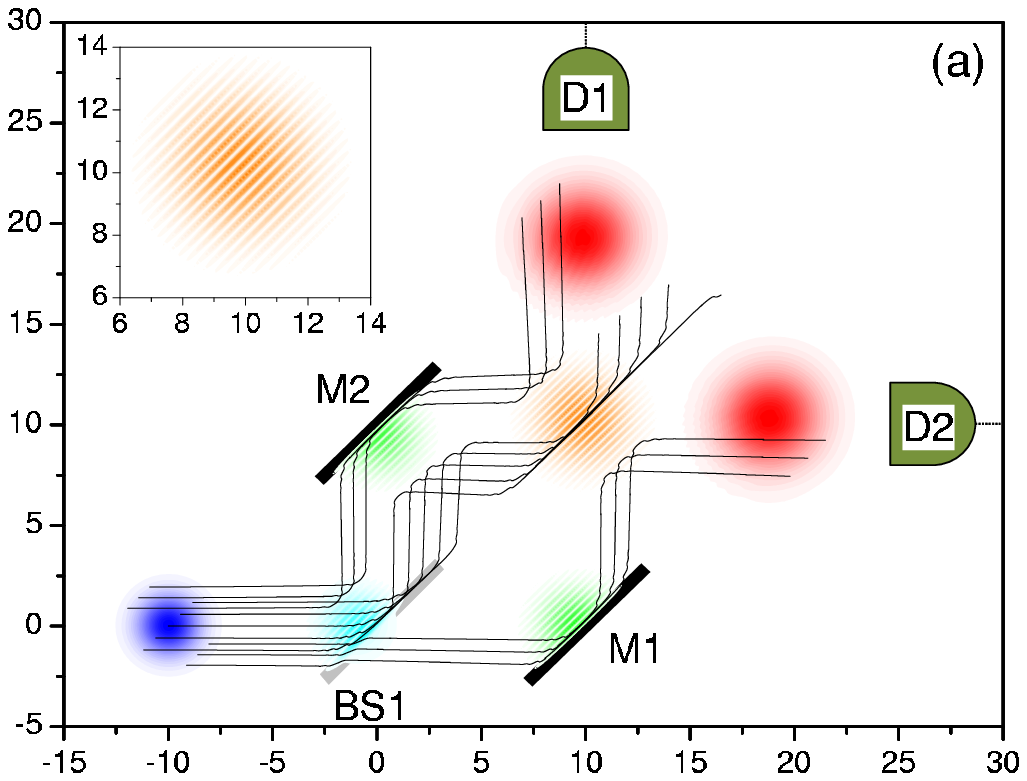}
 \hspace*{4pt}
 \includegraphics[width=6cm]{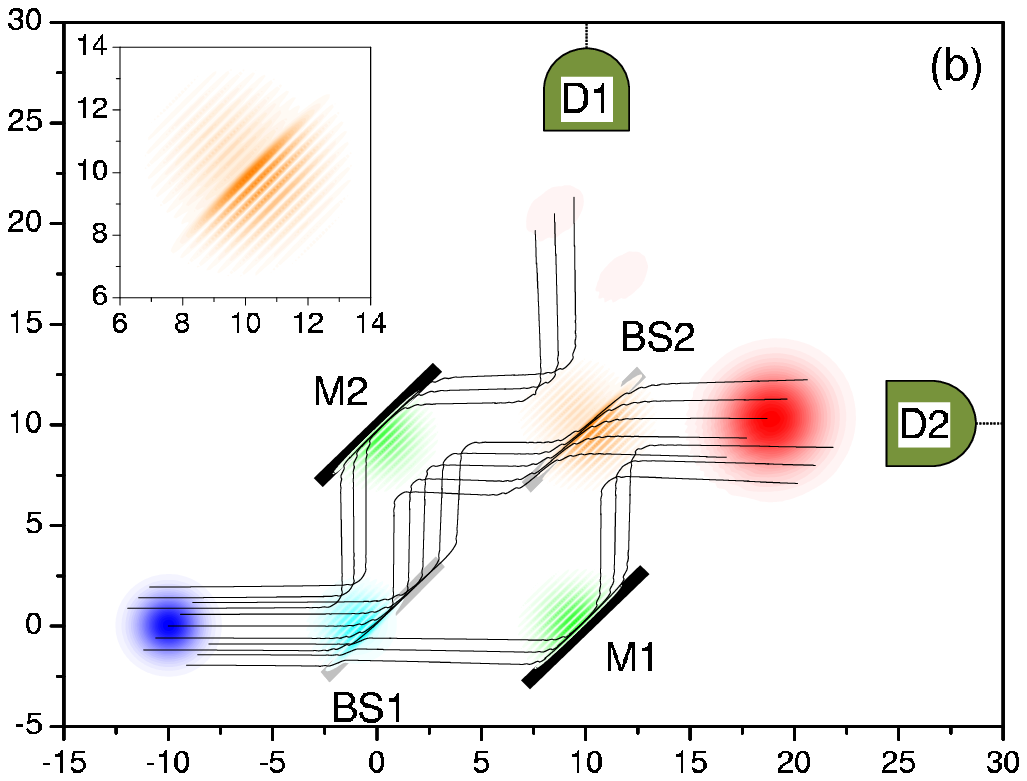}
 \caption{Numerical simulations\cite{sanz-wheeler} of the open (a) and
  closed (b) interferometer configurations involved in Wheeler's delayed
  choice.
  The background monochrome contour-plots correspond to different
  stages in the evolution of the system wave function inside the
  interferometer (see text for details): blue: initial state (Gaussian
  wave packet); light blue: splitting at BS1; green: reflection at the
  mirrors (M1 and M2); orange: superposition of the two wave packets at
  the position where BS2 should be allocated; red: final stage (wave
  packets in their way to the corresponding detectors, D1 and D2).
  In the insets of each panel, a magnification of the probability
  density in the region around BS2.
  The black solid lines represent ensembles of Bohmian trajectories
  starting with initial positions covering the corresponding regions
  of the initial probability density.}
 \label{fig2}
 \end{center}
\end{figure}

Let us now revisit the well-known delayed-choice experiment proposed by
Wheeler\cite{wheeler-bk} in the late 1970s ---a paradigm of the mystery
entailed by quantum mechanics--- to illustrate how Bohmian mechanics
removes the paradoxical aspects introduced by this experiment.
The main idea behind this experiment is very simple: to reveal the
puzzling dual wave-corpuscle nature of quantum systems within the same
experiment.
To this end, Wheeler considers an optical Mach-Zehnder interferometer
with a movable second beam splitter.
Moreover, the experiment is performed in such a way that at each time
there is only one photon inside the interferometer.
For a visual representation of the interferometer configurations
described below, see Fig.~\ref{fig2} (do not confuse here the Bohmian
trajectories displayed with usual optical geometric paths).
When the photon enters the interferometer, the first (fixed) beam
splitter (BS1), oriented at 45$^\circ$ degrees with respect to the
photon incidence direction, may produce direct transmission towards a
mirror M1 with a 50\% of probability, or a perpendicular deflection
(reflection) towards a mirror M2.
In either case, when the photon reaches the mirrors, it gets deflected
90$^\circ$ with respect to the corresponding photon incidence direction.
Eventually, in an open configuration (see Fig.~\ref{fig2}(a)), i.e.,
without the second beam splitter BS2, an arrival can be registered by a
detector D1 if the photon followed the transmitted path (let us denote
it by P1), or by D2 if otherwise it followed the reflected path (P2).
This is a typical corpuscular scenario, describable in terms of
classical optics (the photon follows geometric rays).
On the other hand, in a closed configuration (see Fig.~\ref{fig2}(b)),
when BS2 is introduced at the place where the paths P1 and P2 intersect,
the photon will always be detected by D2.
In this case the photon displays its wave behavior: at BS2 the
components of the associated wave interfere destructively in the
direction of P1 and constructively along P2.
Following the Bohr-Einstein debates,\cite{bohr-einstein} Wheeler
reformulated one of the main questions coming from them: when does the
quantum system make the choice to behave as a wave or as a particle?
Instead of considering a double slit, Wheeler assumed the above
interferometer, but with BS2 movable, so that it could be removed or
inserted once the photon had passed through BS1.
The result is very interesting: regardless of when BS2 is put into
play, the photon always behaves as it should, that is, just as if it
could anticipate what is going to happen in future: the photon makes
a delayed choice.
Although initially conceived as a thought-experiment, this experiment
has already been carried out in the laboratory in many different ways,
but always confirming this challenging dual behavior.\cite{wheeler-exp}

The weirdness of Wheeler's experiment readily dissipates if one starts
describing it properly, that is, using quantum mechanics since the very
beginning, without any aid of classical corpuscle-based model.
A priori the usual Schr\"odinger or Heisenberg descriptions might seem
of little help, since they essentially stress the role of probabilities.
However, if we keep in mind the fact that there is an associated phase
dynamics (and therefore an inherent nonlocal velocity field), things
change.
This is the idea that first Bohm and coworkers,\cite{bohm1985} and
later on Hiley and Callaghan,\cite{hiley2006} tried to convey by
analyzing the experiment in terms of Bohmian mechanics.
Because Bohmian trajectories, which in essence are elements to make
apparent the flow of such probabilities, obey the non-crossing
rule\cite{sanz2008} (see Sec.~\ref{sec22}), what happens is that the
photon always behave in the same way.
If BS2 is absent, because the trajectories coming from P1 and P2 cannot
cross the symmetry line at 45$^\circ$, those coming from P1 are
reflected in the direction of D2, and those from P2 in the direction
of D1.
That is, it is not that the photon follows P1 or P2 until reaching the
corresponding detector, but there is an exchange in the directionality
of the associated quantum flow.
On the other hand, if BS2 is introduced, even once the photon is
inside the interferometer, the wave recombination process taking place
at this beam splitter produces that the two sets of trajectories will
eventually go into one detector, namely D2.
This all-the-way wave behavior (the classical corpuscle notion just
disappears, since it is not necessary at all) is illustrated in
Fig.~\ref{fig2} by means of the numerical simulation\cite{sanz-wheeler}
of the open (a) and closed (b) interferometer configurations described
above.
As it can be seen, there is no choice of the photon at all, but just a
modification of the boundary conditions affecting its wave function,
which naturally gives rise to different outcomes, regardless of whether
BS2 is introduced or removed once the wave function has started its
evolution inside the interferometer.
This kind of realistic simulations are very important to better
understand the physics that is taking place in apparently paradoxical
situations, as it has been recently shown in the case of atomic
Mach-Zehnder interferometry,\cite{sanz2015} for example, which is
typically considered to analyze and discuss fundamental complementarity
issues due to its suitability to this purpose.\cite{pritchard}


\section{Pushing Quantum Mechanics Hard Enough}
\label{sec4}

Quantum mechanics has proven to be the most successful theory ever
devised (at least, to date).
This theory not only has addressed the most fundamental physical
problems, but its applications constitute an important part of our
everyday life (actually, more sophisticated applications are still
to come).
However, the fact quantum phenomena cannot be understood under the same
logic as classical ones brings in a puzzling situation, which has left
open a tough debate on its interpretation since the 1930s.
Probably one may think that this debate will never be really closed (or
at least until we will be able to devise a new, more general theory)
and that, in such a case, it might be pointless to continue talking
about quantum philosophical issues or trying to further develop the
area of the quantum foundations.
Evidently, this leads to a sort of hopeless situation, with a
remarkably close resemblance to Plato's myth of the cave: we are
enforced to perceive the shadows cast on the wall of the cave (our
reality) by the real world (the Reality), without possibility to ever
reaching a true understanding of the physical world.

However, we have seen above that, even within such a harsh scenario,
there is still room enough to further explore the quantum world from a
pragmatist's point of view, just playing around with the quantum rules
and its many way to formulate them.
One only needs to look back for a while and make a reflection on how
the quantum theory has conceptually developed since its early stages,
what can be done at present, and which alternative routes can be
followed.
These are the essential ingredients for new quantum developments and
advances.
Based on recent achievements, the Bohmian formulation of the quantum
theory seems to be one of these routes, which allows us to understand
how quantum systems evolve obeying a non-observable (i.e., not directly
accessible in the experiment) phase dynamics.
Perhaps it will not be possible to determine how a real individual
particle evolves without disturbing it, but at least now we have a
tool to understand how the flow of many of these identical particles
evolves in configuration space.
This has helped us to understand that there are no paradoxes in the
quantum theory, but only a misconception about how quantum systems
behave, anchored in old-fashioned classical prejudices.
In this regard, following Nobel Laureate Anthony
Leggett,\cite{newscientist} ``if we push quantum mechanics hard enough
it will break down and something else will take over ---something we
can't envisage at the moment.''


\section*{Acknowledgments}

The author thanks Roman Ryutin and Vladimir Petrov for their kind
invitation to participate in HEPFT2014, and Anton Godizov for his kind
assistance and friendship during the event.
Support from the Ministerio de Econom{\'\i}a y Competitividad (Spain)
under Project No. FIS2011-29596-C02-01 as well as a ``Ram\'on y Cajal''
Research Fellowship with Ref. RYC-2010-05768 is acknowledged.




\end{document}